\input{aipcheck}

\documentclass[
    ,final            
  ]
  {aipproc}

\layoutstyle{6x9}

\begin{document}

\title{Light Curves of Radio Supernovae}


\author{M. T. Kelley}{
  address={Physics Dept., Marquette University, PO Box 1881, Milwaukee, WI 53201},
 email={mattew.kelley@mu.edu},
}

\author{C. J. Stockdale}{
  address={Physics Dept., Marquette University, PO Box 1881, Milwaukee, WI 53201}
}
\author{K. W. Weiler}{
  address={Naval Research Laboratory}
}
\author{C. L. M. Williams}{
  address={Massachusetts Institute of Technology}
}
\author{N. Panagia}{
  address={Space Telescope Science Institute}
}
\author{R. A. Sramek}{
  address={National Radio Astronomy Observatory}
}
\author{J. M. Marcaide}{
  address={University of Valencia}
}
\author{S. D. Van Dyk}{
  address={Spitzer Science Center, California Institute of Technology}
}

\begin{abstract}
We present the results from the on-going radio monitoring of recent type II supernovae (SNe), including SNe 2004et, 2004dj, 2002hh, 2001em, and 2001gd. Using the Very Large Array\footnote{The VLA telescope of the National Radio Astronomy Observatory is operated by the Associated Universities, Inc. under a cooperative agreement with the National Science Foundation.} to monitor these supernovae, we present their radio light-curves. From these data we are able to discuss parameterizations and modeling and make predictions of the nature of the progenitors based on previous research. Derived mass loss rates assume wind-established circumstellar medium, shock velocity $\sim 10,000$ km s$^{-1}$, wind velocity $\sim 10$ km s$^{-1}$, and CSM Temperature $\sim 10,000$ K. 
\end{abstract}

\maketitle


\section{Radio Light Curves}

Over the course of the last six years, we have made radio observations of five recent supernovae (SNe), SN~2001gd (type IIb; NGC~5033), SN~2001em (type IIn; UGC~11794), SN~2002hh (type II; NGC~6946), SN~2004dj (type IIP; NGC~2403), and SN~2004et (type IIP; NGC~6946). In this paper, we present the radio light curves of these SNe.  The parameterized model fits of the data using the model presented by Sramek et al. (2005) and a discussion of these results can be found in the proceedings of Stockdale et al. (2007).

\begin{figure}
  \rotatebox{270}{\includegraphics[height=.3\textheight]{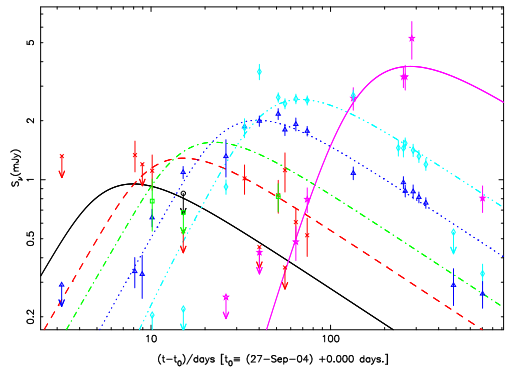}}
  \caption{The radio light curve of the Type IIP SN~2004et.}
\end{figure}
\begin{figure}
  \rotatebox{270}{\includegraphics[height=.3\textheight]{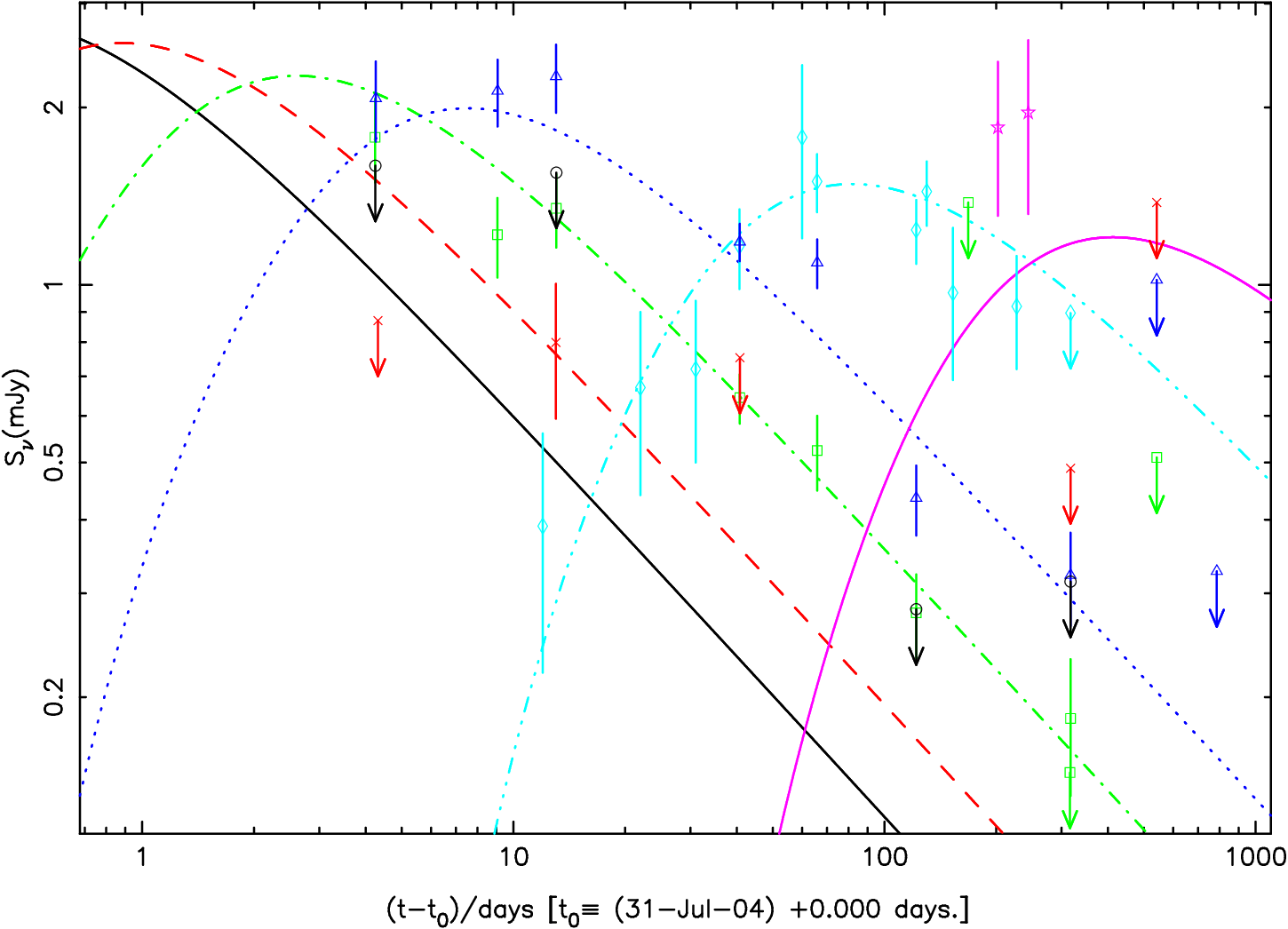}}
  \caption{The radio light curve of the Type IIP SN~2004dj.}
\end{figure}
\begin{figure}
  \rotatebox{270}{\includegraphics[height=.3\textheight]{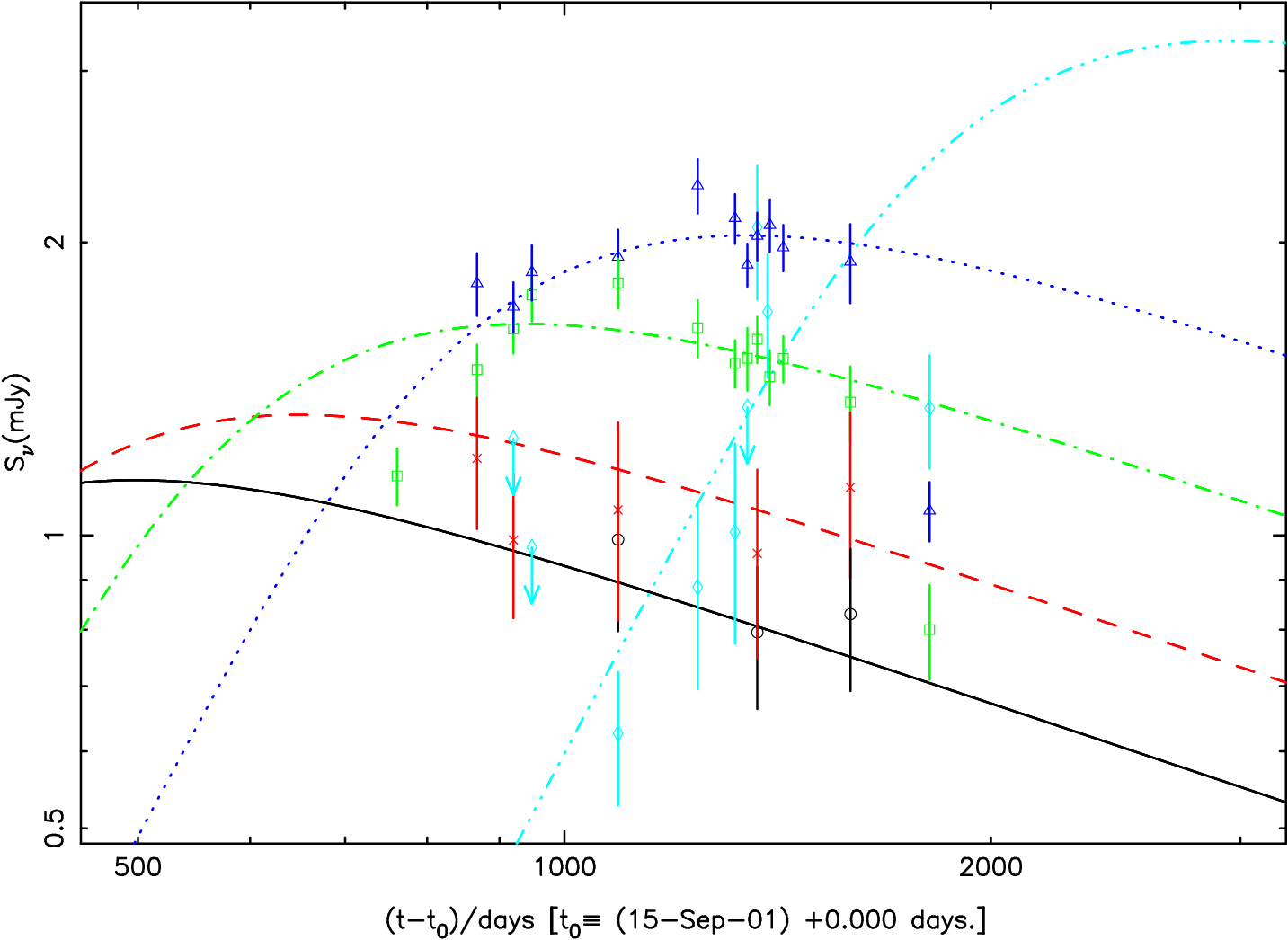}}
  \caption{The radio light curve of the Type IIn SN~2001em.}
\end{figure}
\begin{figure}
  \rotatebox{270}{\includegraphics[height=.3\textheight]{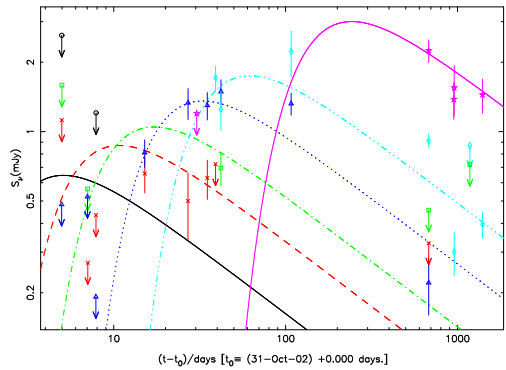}}
  \caption{The radio light curve of the Type II SN~2002hh.}
\end{figure}
\begin{figure}
  \rotatebox{270}{\includegraphics[height=.3\textheight]{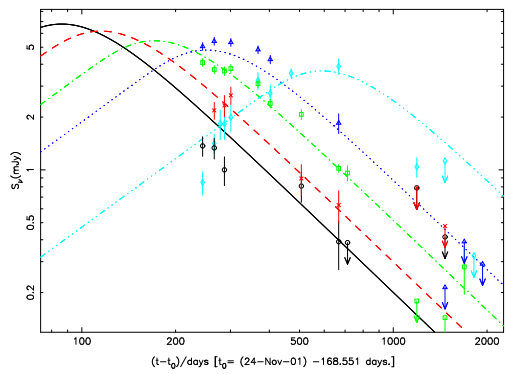}}
  \caption{The radio light curve of the Type IIb SN~2001gd.}
\end{figure}

\section{Results}
Our models indicate that the type IIb SNe thicker and denser circumstellar medium (CSM) than the type IIP SNe in our sample. This is illustrated by the light curves peaking at later times. This supports the idea that type IIb SNe are a transitional class to type Ib/c SNe which are thought to be binary massive star systems. Our light curves also indicate a more complex interaction with the CSM and the SN blastwave for the type IIb SNe, possibly resulting from a large, rapid mass loss of the progenitor star. The light curve of SN~2001em indicates that it is encountering a dense, thick CSM which is in stark contrast to typical Ib/c SNe which are not usually strong radio emitters. Further VLA and VLBI studies may provide useful insight into this unusual SN.  

\begin{theacknowledgments}
MTK was supported by the NASA Wisconsin Space Grant Consortium.  CJS is a Cottrell Scholar, supported by Research Corporation.  KWW thanks the Office of Naval Research for the 6.1 funding which supports his research.
\end{theacknowledgments}


\begin{thebibliography}{9}
\bibitem{Stockdale}
Stockdale, C.J.; Kelley, M.T.; Weiler, K.W.; Panagia, N.; Sramek, R.A.; Marcaide, J.M.; Williams, C.L.M.; Van Dyk, S.D.; "Recent Type II Radio Supernovae" in \emph{SN~1987A: 20 Years After - Supernovae and Gamma-Ray Bursters} ed. by Weiler, K. W., and Immler, S., American Institute of Physics, U.S.A., 2007.
\bibitem{Sramek}
Sramek R.A.; Weiler K.W.; Panagia N.; "Radio Supernovae" in \emph{Cosmic Explosions On the 10th Anniversary of SN 1993J}, ed. by Marcaide, J.M., and Weiler, K.W., Springer-Varlag Berlin Heidleberg, Germany, 2005.
\end{thebibliography}
\end{document}